\documentclass[acmsmall,screen]{acmart}
\usepackage{booktabs} 
\usepackage{textcomp} 
\usepackage{tabularx, array, makecell, float, multirow}

\AtBeginDocument{%
  }

\author{Kexin Bella Yang}
\authornote{Both authors contributed equally to this research. \textbf{Corresponding author.}}
\affiliation{%
  \institution{Human-Computer Interaction Institute, Carnegie Mellon University}
  \city{Pittsburgh}
  \country{USA}
}
\email{kexiny@andrew.cmu.edu}

\author{Menghan Liu}
\authornotemark[1]
\affiliation{%
  \institution{University of Washington}
  \city{Seattle}
  \country{USA}
}
\email{menghanl@uw.edu}

\author{Liyi Xu}
\authornotemark[1]
\affiliation{%
  \institution{Human-Computer Interaction Institute, Carnegie Mellon University}
  \city{Pittsburgh}
  \country{USA}
}
\email{zczlky4@ucl.ac.uk}

\author{Nikol Rummel}
\affiliation{%
  \institution{Institute of Educational Research, Ruhr-Universität Bochum}
  \city{Bochum}
  \country{Germany}
}
\email{nikol.rummel@rub.de}
 
\author{Vincent Aleven}
\authornote{Corresponding author.}
\affiliation{%
  \institution{Human-Computer Interaction Institute, Carnegie Mellon University}
  \city{Pittsburgh}
  \country{USA}
}
\email{aleven@cs.cmu.edu}

\setcopyright{none}
\renewcommand\footnotetextcopyrightpermission[1]{}
\usepackage{siunitx}

\begin{document}

\title{Balancing Teacher and Student Agency: Co-Orchestration Tool Design Supporting Real-Time Dynamic Pairing}

\begin{abstract}
In human-AI interaction, respecting user agency is essential for fostering trust and sustaining effective use of technology. In educational settings, dynamically integrating individual and collaborative learning offers pedagogical value by supporting personalized, self-paced learning experiences. Prior research has demonstrated the feasibility of this approach through intelligent tutoring systems and human–AI co-orchestration tools. However, how to balance teacher and student control in this process remains largely unexplored. This work explores the design space of how control can be distributed between teachers and students across the orchestration process, using participatory speed dating and a mixed-method analysis. We focus on three stages of the pairing process: before, during, and after, taking context in designing classroom orchestration tools that support teachers in dynamically coordinating student transitions between individual practice and collaborative problem-solving. It contributes empirical insights to the fields of educational technology and HCI by framing these findings within a theoretical design space, emphasizing the balance of multi-stakeholder agency and control. We propose design recommendations for achieving hybrid-control in analytic-based orchestration tools in pairing contexts. We recommend ensuring structured teacher guidance in the beginning, while progressively increasing student autonomy over time as activities unfold.
\end{abstract}




\maketitle
\renewcommand{\textcolor}[2]{#2}

\section{Introduction}
In Human-Computer Interaction (HCI), agency refers to the control and influence users have when interacting with systems, including the ability to make choices, perform actions, and perceive their impact on the system \cite{Bennett2023-jg,Limerick2014-ie,Zacarias2012-cp}. There is a consensus on the value of autonomy and user agency for diverse communities within HCI and CSCW \cite{Bennett2023-jg,murray2024agency,o1992agency}. User agency may help align AI recommendations with user preferences and foster transparency and user trust in AI systems \cite{Amershi2019-jk,Binns2018-em,Shneiderman2003-ey, Hollnagel2005-uo}. In human-AI interaction, designing for multiple stakeholders (students, teachers, administrators) control requires flexible, adaptive systems with nuanced design considerations \cite{Holstein2019-io}. Additionally, the different roles stakeholders take may influence their desired agency. For example, depending on teachers' preferred level of engagement with generative AI system, they may want to take the role of Observer, Adopter, Collaborator, Innovator \cite{Zhai2024-ac}.\textit{Classroom orchestration} involves planning, monitoring, and coordinating multiple complex classroom activities in real time \cite{Dillenbourg2010-do,Holstein2019-gi}. In recent years, there has been increasing emphasis on leveraging analytic-based classroom orchestration in education and CSCL domain \cite{Das2023-tu,tissenbaum2019supporting}, which involves a collaborative effort between teachers and educational technology systems to facilitate and manage classroom tasks \cite{Yang2023-lq,Holstein2019-gi,Olsen2020-jg}. 

In various education levels (e.g., K-12, post-secondary \cite{yildiz2023institutional,murray2024agency}), students commonly engage in combinations of individual or collaborative learning activities \cite{Diziol2007-xe,Van_Leeuwen2019-gz}, which can be more effective than either mode alone \cite{Olsen2019-ny}. Traditionally, teachers plan in advance how students switch between individual and collaborative activities. \textit{Dynamic combinations of individual and collaborative learning}, sometimes called “dynamic pairing” may happen opportunistically determined by real-time learning situations of the students \cite{Echeverria2020-si,Yang2023-lq, Yang2021-np,Yang2021-aa}. Such dynamic combinations may not occur all at the same time for everyone in the class, and students may work on personalized content that best serves their own learning needs \cite{Olsen2020-jg,Yang2023-lq}. A field study confirmed the feasibility and indicated the potential educational benefits of such dynamic combinations \cite{Yang2023-lq}.
    
\begin{figure}
    \centering
    \includegraphics[width=0.75\linewidth]{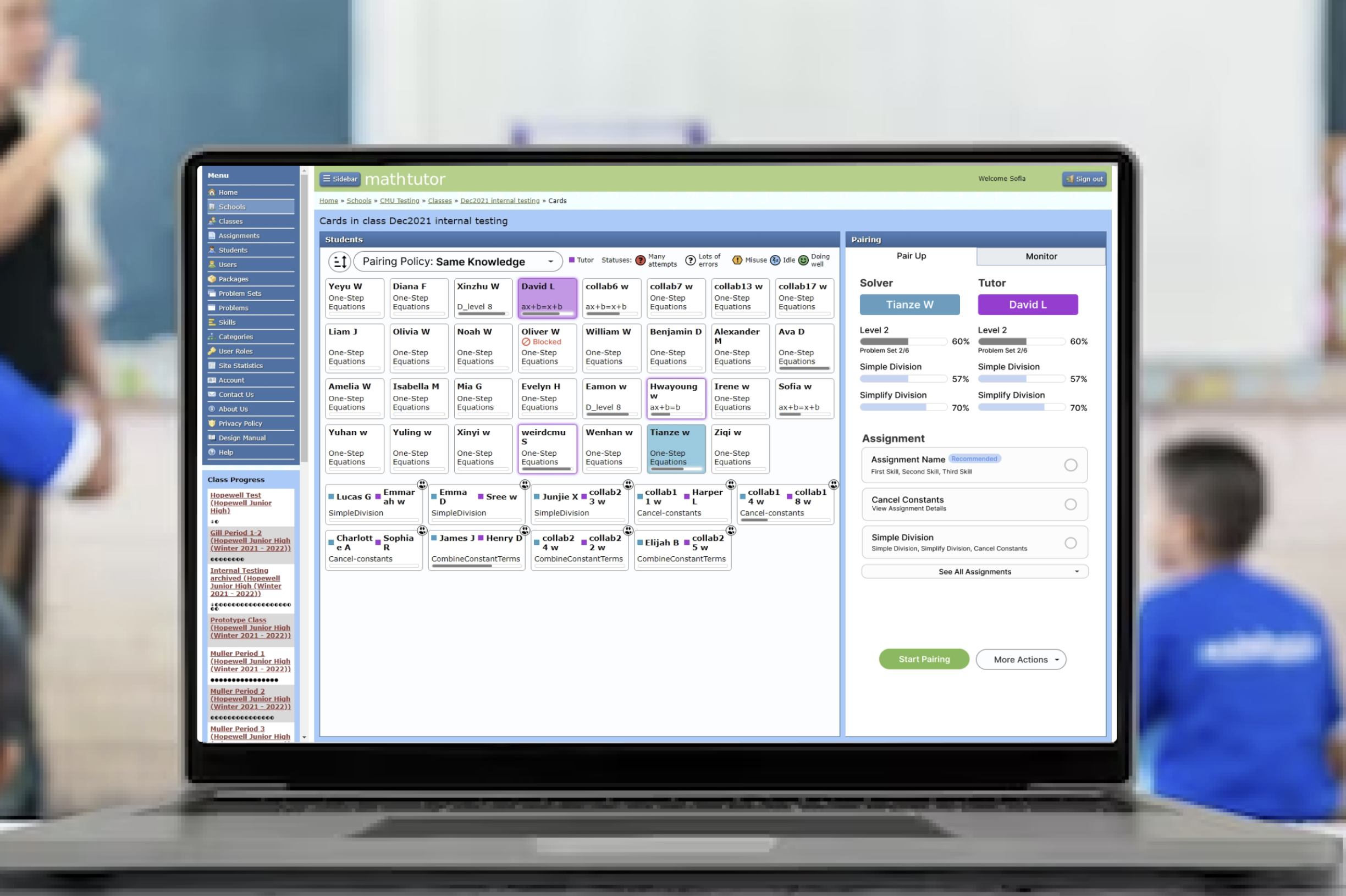}
    \caption{Mockup of classroom use for the teacher-facing human-AI co-orchestration tool for dynamic combinations of individual and collaborative learning, where teachers can 1) view students real-time learning analytics coming from the students' AI-based tutoring software, 2) receive AI system’s intelligent pairing suggestions based on students’ mastery of knowledge components targeted in the work with the tutoring software, 3) dynamically orchestrate activities in class including deciding the learning tasks, mode, partner, and timing}
    \label{Fig.1.}
\end{figure}

Real-time dynamic pairing involves frequent educational decisions (including pairing/unpairing timing, partners and content) \cite{Echeverria2020-si,Yang2023-lq,Yang2021-np,Yang2021-aa}. When teachers decide solely, students may feel a lack of agency in the learning process \cite{Yang2023-lq,han2025helping}. While teachers  respect the’ student's agency and control, they can be unsure how to best allow student input in the pairing process when interacting with the tool (Fig.\ref{Fig.1.})). Studies have explored hybrid control options between students, teachers, and AI (e.g. \cite{Echeverria2020-si,Yang2021-np}), but little is known about which decisions stakeholders wish to share, how they prefer sharing, and where views align or differ.
In our study, individual learning refers to students solving mathematics problems within an intelligent tutoring system (ITS) \cite{Long2015-sp, Holstein2019-io}. Collaborative learning takes the form of peer tutoring within the ITS, in which one student assumes the role of solver and works through the problem, while the other acts as tutor by providing hints and corrective feedback \cite{Walker2014-bm, Long2013-cn, Nagashima2023-va}.The orchestration tool is a teacher-facing support system that enables instructors to dynamically pair and unpair students based on real-time learning analytics, including students’ learning mastery (tracked via Bayesian Knowledge Tracing) and metacognitive states (see Section 3.1 for details) \cite{Yang2022-nj}. The AI components of our educational system primarily consist of (1) a knowledge-tracing algorithm that models students’ mastery levels within the individual ITS, and (2) rule-based mechanisms that generate pairing and task assignment suggestions based on students’ knowledge distance and struggle status, as inferred from real-time interaction logs and detector outputs from the tutoring software. Leveraging these AI components, students receive intelligent problem selection—namely, personalized problem sets tailored to their mastery of specific equation-solving skills—while teachers are provided with real-time recommendations for pairing partners and corresponding collaborative assignments. 
    
The theoretical motivation is to address a literature gap by contributing empirical insights within a structured design space, on balancing hybrid control and multi-stakeholder agency in classroom orchestration. Specifically, our study aims to investigate the preferences of teachers’ and students’ in hybrid-control (RQ1), and inform the design of future teacher-facing orchestration tools in pairing and collaborative learning settings (RQ2). We used a participatory speed dating (PSD) method (well-leveraged in CSCW and HCI research literature \cite{Kinnula2023-pf,Zhang2023-tv}, with 17 teachers and 13 students to efficiently understand their perspectives. We performed a mixed-method analysis, by triangulating the quantitative (ranking) and qualitative data (reasoning) to inform a comprehensive understanding of multi-stakeholder preferences.

\section{RELATED WORK AND THEORETICAL BACKGROUND}
This study builds on previous research on user agency in HCI and education, human-AI co-orchestration, and dynamic combinations of individual and collaborative learning.

\subsection{Supporting stakeholders’ agency and control in education settings}
Teachers and learners constitute two distinct stakeholder groups with fundamentally different needs, responsibilities, and preferences in educational settings \cite{Frosig2024-jo}. In the learning sciences and AI-based educational technology, control and agency refer to the extent to which teachers and students can make decisions, regulate their own behavior \cite{Edward2000-hj,Reeve2006-rs}, and influence teaching and learning practices and outcomes \cite{Hooshyar2023-ve}. Agency plays a critical role in shaping teachers’ trust in AI-enabled educational systems \cite{Frosig2024-jo,Tanya-Nazaretsky-Moriah-Ariely-Mutlu-Cukurova-Giora-AlexandronUnknown-qx}, as well as in addressing diverse student needs and classroom heterogeneity \cite{Shavelson1981-lo}. However, an overabundance of control options may hinder efficiency and disrupt the fluidity required for real-time classroom orchestration \cite{Holstein2021-fp,Yang2023-lq,Holstein2019-gi}.

Prior research has extensively examined learner agency and engagement, particularly through mechanisms such as learner control \cite{Long2015-sp} or shared control \cite{Corbalan2008-md,Corbalan2006-gj} over learning tasks and problem selection \cite{Frosig2024-jo,Nguyen2018-in,Long2015-sp,Corbalan2008-md,Liuqing-Chen-Zhaojun-Jiang-Duowei-Xia-Zebin-Cai-Lingyun-Sun-Peter-Childs-and-Haoyu-ZuoUnknown-vh}. For example, Corbalan et al. demonstrated that shared control between learners and systems in task selection increased task involvement and effort, ultimately leading to improved learning outcomes \cite{Corbalan2008-md}. Consistent with this line of work, student agency has been associated with higher motivation and engagement, as well as deeper learning and conceptual understanding \cite{Taub2020-im,Dimitra_Tsovaltzi2013-ol,Robertson2017-st,han2025helping,Cook-Sather2014-wt}. At the same time, excessive learner control can overwhelm students and hinder effective self-regulation \cite{Long2015-sp,Bjork2013-bc}, particularly for novice learners who may lack the expertise needed to make informed learning decisions.

Taken together, these findings suggest that neither full learner control nor full instructor control is universally desirable. Instead, prior work has argued for balanced approaches that combine structured teacher guidance with opportunities for learner autonomy \cite{Clark2012-zd}. Yet, sharing control between instructors and learners, if not carefully designed, can increase instructors’ workload, as learners are often still developing the skills required to effectively assume these responsibilities \cite{Holstein2021-fp}. A theoretical framework proposed by Eshel and Kohavi provides insights into how classroom control can be shared between teachers and students. This framework conceptualizes control along two relatively independent dimensions: (1) student control and (2) opportunities for self-directed learning \cite{Eshel2003-vv}. Rather than treating teacher and student control as a zero-sum tradeoff, Eshel and Kohavi argue that both groups can simultaneously exercise high levels of control through hybrid control, in which teachers retain authority over structuring and guiding the learning process, while students are afforded opportunities for self-directed learning, making choices, and contributing to decision-making \cite{Eshel2003-vv}. Within AI-enhanced classrooms, several frameworks have further examined how control and agency can be distributed across human and technological actors. Molenaar proposed six levels of control distribution between teachers and technology, ranging from teacher-only control to full automation \cite{Molenaar2022-to}. Complementarily, Yang et al. examined how teachers’ decision-making processes—such as proposing, evaluating, and deciding collaborative pairing choices—can be shared among teachers, students, and AI systems \cite{Yang2021-np}.

While these frameworks primarily focus on teacher–AI or system-level distributions of agency \cite{Molenaar2019-do,Echeverria2020-si}, important gaps remain. In particular, little is known about how hybrid control should be systematically designed and distributed across different stages of educational technology tools that orchestrate collaborative learning, or how teachers and students perceive these hybrid control design choices. Addressing these gaps, the current study contributes a design space that captures fine-grained, multi-stakeholder perspectives on control distribution in human–AI co-orchestration. Our findings offer empirically grounded insights into how classroom orchestration systems for collaborative learning can support decision-making that balances agency, accountability, and orchestration efficiency.

\subsection{Co-orchestration and dynamic combination of individual and collaborative learning}
Traditional classroom orchestration requires teachers to manage complex, interdependent activities in real time \cite{Dillenbourg2010-do,Holstein2019-gi}. Recently, research has increasingly focused on human–AI co-orchestration \cite{Yang2023-lq,Holstein2019-gi,Olsen2020-jg}, which leverages the complementary strengths of teachers, learners, and AI systems to support classroom management rather than placing the full orchestration burden on instructors \cite{Sharples2013-rn}. Co-orchestration may distribute responsibilities among (1) instructors and learners, (2) instructors and AI-driven instructional agents, or (3) instructors, learners, and AI agents \cite{Holstein2021-fp}. This distribution can take the form of \textit{role splitting}, in which humans and AI assume distinct orchestration tasks, or \textit{role sharing}, in which both contribute to the same tasks \cite{Holstein2021-fp}. Holstein and Olsen \cite{Holstein2021-fp} further articulate a vision of shared orchestration among instructors, learners, and AI agents, integrating both instructor–AI and instructor–learner collaboration.

Human–AI co-orchestration tools often provide real-time learning analytics that help teachers monitor student progress, provide timely interventions, and make informed instructional decisions \cite{Yang2023-lq,Lawrence2024-mf,Olsen2020-jg}. For example, the FACT system alerts teachers when students need support and suggests whom to assist and how \cite{VanLehn2018-bv}. Previous work demonstrates that such tools hold substantial promise in enabling adaptive instruction and supporting the management of dynamic classroom contexts (cf. \cite{Manathunga2015-ih,Prieto2018-dy,Manathunga2019-me,Dillenbourg_undated-mt,Yang2021-aa,Holstein2019-gi, Yang2023-lq}).

This study focuses on orchestration tools that support dynamic transitions between individual and collaborative learning, two widely used instructional modes with complementary benefits. Collaborative learning fosters mutual knowledge construction, whereas individual learning supports processes such as induction and refinement (cf. \cite{Dimitra_Tsovaltzi2013-ol,Etelapelto2013-eg}). Combining these approaches can be more effective than either alone; for instance, prior work shows that students make fewer errors and request fewer hints when instruction integrates both modes \cite{Olsen2019-ny}. In personalized learning environments, dynamically switching between learning modes enables instruction to better align with individual student needs \cite{Olsen2020-jg,Echeverria2020-si,Yang2021-aa,Yang2021-np}. Accordingly, technology ecosystems have been developed that allow students to transition between learning modes at moments teachers deem most pedagogically beneficial \cite{Yang2022-nj,Yang2023-lq}.

Prior research reveals tensions in teacher and student agency in orchestration tool design \cite{Yang2023-lq}: students desire greater autonomy in partner selection, while teachers prioritize classroom order and instructional effectiveness. These tensions inform the need to carefully balance control and agency in the design of co-orchestration systems. Relatedly, in a classroom Wizard-of-Oz technology probe, Echeverria et al. simulated hybrid pairing control among students, teachers, and AI, highlighting the importance of shared agency during transitions between learning modes \cite{Echeverria2020-si}. However, this work did not address how such hybrid control might be implemented in practice—an open challenge that this study seeks to address.
    
\section{MOTIVATION, RESEARCH QUESTIONS AND STUDY CONTEXT}
\subsection{Research question and design motivation}
Building on prior work, this study is motivated by recurring challenges in classroom orchestration, including determining when students are ready to transition \cite{Echeverria2020-si, Yang2023-lq}, balancing teacher authority with student preferences in partner and role assignment \cite{Echeverria2020-si, Yang2023-lq, Holstein2021-fp}, selecting collaboration tasks aligned with learners’ knowledge levels \cite{Bjork2013-bc, Corbalan2008-md}, monitoring collaboration quality and productivity in real time \cite{Holstein2019-gi, Olsen2020-jg}, and supporting smooth transitions without disrupting learning continuity \cite{Yang2023-lq, Echeverria2020-si}. These challenges are formalized in Section 4 and directly inform our design motivations: supporting appropriate student autonomy in pairing, preserving teacher autonomy for informed orchestration decisions, and balancing control between teachers and students to enable smooth and effective collaboration. While the design primarily focuses on teacher–student control balance, it also incorporates limited elements of system control where appropriate. We focus on the following research questions:
    
	\textbf{RQ1}: How do teachers and students prefer to distribute control and agency before, during and after the dynamic pairing process?
    
	\textbf{RQ2}: How might we derive design implications for balancing teacher-student agency, for future analytic-based classroom orchestration tools (especially in pairing and collaborative learning settings)?

\subsection{Study context: technology ecosystem for orchestrating dynamic pairing}

\begin{figure}[htbp]
  \centering
  \includegraphics[width=1\linewidth]{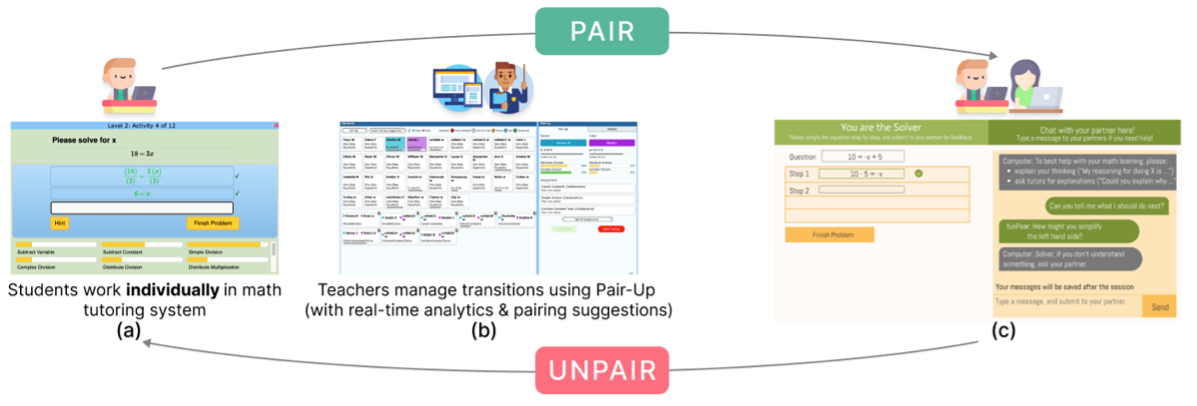}  % 
  \caption{Co-orchestration technology ecosystem for dynamic combinations of individual and collaborative learning: intelligent tutoring systems for learning linear algebra individually (a) and collaboratively (c), and a teacher-facing co-orchestration tool (b)}
  \label{Fig.2.}
\end{figure}

The technology ecosystem (Fig.\ref{Fig.2.}) in this study consists of two intelligent tutoring systems and a teacher-facing co-orchestration tool that allows dynamic pairing \cite{Yang2023-lq,Yang2022-nj}. The AI-based tutoring system involves Bayesian Knowledge Tracing, a probabilistic method used in Intelligent Tutoring Systems, to track students' knowledge level of algebra and equation solving over time based on their correct and incorrect responses \cite{yudelson2013individualized}. The collaborative learning system is a peer-tutoring system, where one student takes on the role of the Tutor, who guides the Solver to solve algebra problems \cite{Walker2014-bm}. It provides the Tutor with feedback on their tutoring and math hints, including features for marking steps as correct or incorrect, providing hints, and facilitating communication between the Tutor and the Solver through a chat module. The orchestration tool helps teachers make real-time, data-informed decisions about when students should switch between learning modes (through rule-based pairing suggestions based on BKT, knowledge distance, and student struggle situation). It also suggests to teachers the recommended collaborative learning assignment based on students' personalized mastery levels on knowledge components  \cite{Lawrence2024-mf,Olsen2020-jg,Yang2023-lq}.
\section{METHODS}
In CSCW literature, researchers leveraged participatory design and stakeholder-centered design in various applications \cite{Zhang2023-tv,Kinnula2023-pf}. We used a Participatory Speed Dating (PSD) approach \cite{Holstein2019-gi,Muller_and_Sarah_Kuhn_undated-gj}, a design research method used to rapidly gather feedback and generate new ideas using storyboarding \cite{Holstein2019-gi,Khai_N_Truong_Gillian_R_Hayes_and_Gregory_D_Abowd_undated-im}, to understand the perspectives of both teachers and students. Adapted from participatory design \cite{Spinuzzi2005-en}, the Participatory Speed Dating method provided an opportunity to involve multiple stakeholders early in the ideation phase for a general exploration, ensuring that the design meets the needs of users while enhancing user satisfaction \cite{Holstein2019-gi,Muller_and_Sarah_Kuhn_undated-gj,Steen2013-lt}. Moreover, the use of speed dating facilitates rapid exploration and validation of design concepts, enabling quick prioritization of user needs and uncovering new design opportunities through structured comparisons and user feedback \cite{Davidoff2007-pb}.

PSD method was particularly suitable to convey complicated design concepts and obtain constructive feedback quickly. Storyboarding (\cite{Holstein2019-gi,Khai_N_Truong_Gillian_R_Hayes_and_Gregory_D_Abowd_undated-im}), allowed us to illustrate various scenarios involving different levels of teacher and student control during the entire pairing co-orchestration process. PSD allowed us to: 1) generate innovative ideas with classroom stakeholders; 2) test large amounts of ideas for different autonomy levels through rapid explorations; and 3) learn both perspectives of teachers’ and students’ effectively. There were two study phases. Phase 1 was to co-generate design ideas for the seven common challenges by involving teachers and students, reducing the potential bias of researchers, and obtaining more user-centered views and stakeholder input. In Phase 2 we uncovered the preferred control levels of teachers and students through rapid evaluation of ideas. By combining users’ rankings of the design ideas with their accompanying verbalized reasoning, we were able to gain deeper insight into their attitudes toward classroom control dynamics and underlying preferences.

\subsection{Study Preparation:  identifying design challenges in different stages orchestrating dynamic pairing}
We began by reviewing prior work on classroom orchestration, the dynamic integration of individual and collaborative learning, and classroom autonomy \cite{Echeverria2020-si,Eshel2003-vv,Holstein2019-gi,Olsen2020-jg,Yang2021-np}. While existing research has examined preparation before team formation (e.g., community-wide deliberation \cite{Wen2017-wt}) and emphasized the importance of clarifying classroom rules and expectations \cite{Stefanou2004-wk}, it remains unclear how students can be adequately prepared for collaboration during ongoing classroom activities. Moreover, teacher-initiated collaboration may overlook students’ readiness and preferred timing, potentially disrupting prior learning. When teachers solely determine partners, roles, and tasks, students may be assigned unwanted partners, unsuitable roles, or content misaligned with their knowledge and learning context, limiting autonomy and engagement. Once collaboration begins, teachers also struggle to monitor engagement and collaboration quality in real time, ensure smooth transitions between individual and collaborative work, and evaluate collaboration outcomes to inform future pairings.

Based on these challenges, we first identified the user journeys of teachers and students in the dynamic pairing process. We organized this process into three stages: \textbf{Stage 1—Before Pairing Up}, \textbf{Stage 2—During Pairing Up} (pairing, collaboration, and unpairing), and \textbf{Stage 3—After Pairing Up}. Across these stages, we identified seven hybrid-control design challenges, each formulated as a “How Might We” questions (see Table.~\ref{tab:hmw_table}).

\begin{table}[h!]
\centering
% This command increases vertical padding (1.5 is 50% more space)
\renewcommand{\arraystretch}{1.5} 

\begin{tabularx}{\columnwidth}{p{4.2cm} X} 
\hline
\textbf{Stages} & \textbf{HMW question} \\
\hline

% --- Stage 1 ---
Stage 1: Before Pairing Up &
Challenge 1 (C1): How might we design a tool to help teachers prepare for the upcoming collaboration process before class? \\
\hline

% --- Stage 2 ---
% Note: If "Stage 2" looks too high, increase the '4' (e.g., to 6 or 8) 
% because multirow counts *lines of text*, not just table rows.
\multirow{4}{=}{Stage 2: During Pairing Up} &
Challenge 2 (C2): How can we design the tool to ensure that collaboration begins at an appropriate time? \\
\cline{2-2}

 &
Challenge 3 (C3): How might we design the tool to assign suitable roles and partners? \\
\cline{2-2}

 &
Challenge 4 (C4): How might we design the tool to select suitable collaboration content for students to work on? \\
\cline{2-2}

 &
Challenge 5 (C5): How might we design the tool to ensure the collaboration is productive? \\
\hline

% --- Stage 3 ---
\multirow{2}{=}{Stage 3: After Pairing Up} &
Challenge 6 (C6): How might we design the tool to ensure students have a smooth transition between collaborative and individual activities? \\
\cline{2-2}

 &
Challenge 7 (C7): How might we design the dynamic co-orchestration tool to effectively evaluate collaboration activity, ensuring good pairing to occur again in the future? \\
\hline

\end{tabularx}
\caption{Seven “How Might We” questions mapped to stages of the dynamic pairing process}
\label{tab:hmw_table}
\end{table}

\subsection{Participants}

We recruited 30 participants in total (17 teachers and 13 students). Phase 1 included 2 teachers and 3 students, while Phase 2 included 15 teachers and 10 students through network connections, snowballing, and social media. Participants completed a one-hour study via Zoom, and as compensation, received Amazon gift cards in the amount of \$15 (students) and \$30 (teachers). The detailed demographics are provided in the Appendix. Before the study, all participants signed a consent form approved by our institution’s IRB (protocol omitted for blind review). 

\subsection{Study Procedure}

\subsubsection{Phase 1: Idea Co-Generation with Teachers and Students.}
\begin{figure}[H]
    \centering
    \includegraphics[width=\textwidth]{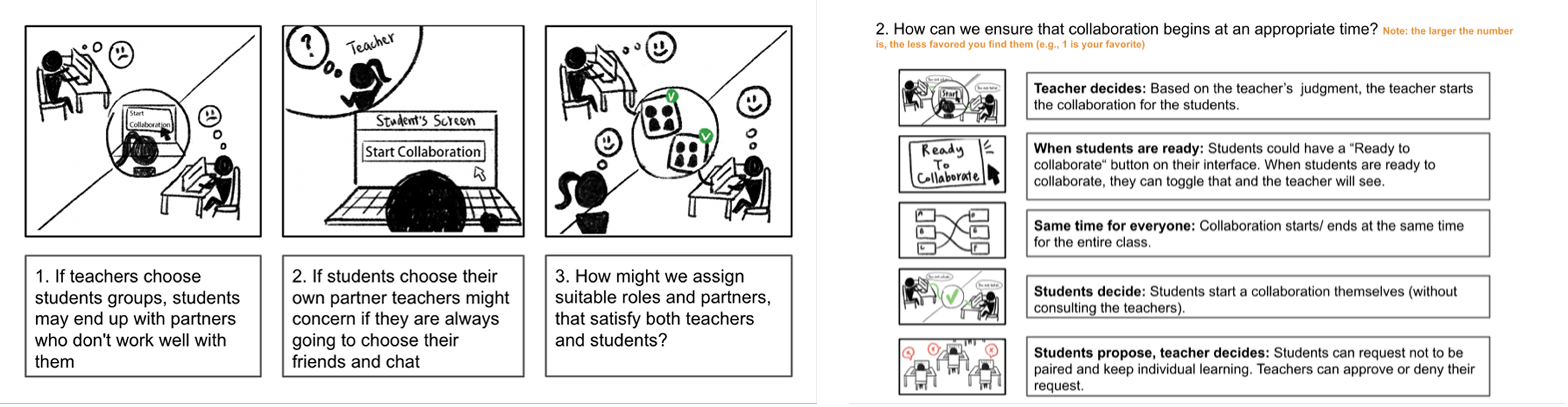}
    \caption{ Left: Example storyboard for Phase 1: Idea co-generation for Challenge 3: collaboration partner; Right: Example storyboard for Phase 2: Idea Evaluation for Challenge 2: Pairing Timing.}
    \label{fig:Group_1000001007}
\end{figure}

We conducted five individual co-generation sessions with 2 teachers and 3 students. After introducing the study and co-orchestration tool functions, participants viewed seven storyboards (e.g., Fig.  \ref{fig:Group_1000001007}), each representing a design challenge. \textcolor{blue} {The researchers explained each challenge and the participants shared initial reactions. We asked participants to ideate possible design solutions to the challenges through concepts or storyboard for at least three challenges, guided by some sample seed designs shown as inspirations, based on prior work \cite{Echeverria2020-si,Yang2023-lq,Yang2021-np}. We did not require participants to limit every brainstormed design ideas to have AI component.} Phase 1 aimed to support generative co-design rather than produce an exhaustive set of ideas, and the sessions reached clear qualitative saturation in the range of concepts generated. The researchers facilitated the process by illustrating ideas using Google Slides.
After combining ideas generated by stakeholders and researchers, across 7 challenges, there was a total of 74 design ideas created (27 researcher-generated, 25 teacher-generated, 22 student-generated). Researchers iteratively refined, voted, and selected ideas to include ideas representing varied levels of hybrid-control for each challenge. 4 to 7 representative solutions for each challenge were selected based on their control level, quality, and popularity, resulting in a total of 34 design ideas, among which, 17 were participant-generated (teacher or student) and 17 were researcher-generated. We illustrated all these design ideas using storyboards, accompanied by 1-2 sentence text explanations (Fig. \ref{fig:Group_1000001007}, right).

\subsubsection{Phase 2: Idea Evaluation by Teachers and Students.}
We conducted 25 1-hour idea evaluation sessions with 10 students and 15 teachers. After introducing the study’s context, participants were shown the 34 design ideas for all 7 challenges, one by one. In each design, they were asked to rank the design idea in each challenge from their most favorite (ranking = 1) to least favorite (ranking = number of design ideas in that challenge). Participants were asked to explain their reasoning for their rankings and choices.
\subsection{Data Collection and Analysis}
We performed a mixed-method analysis, triangulating \textit{quantitative} rankings of design ideas with \textit{qualitative} insights from participants’ verbal reasoning.

	For qualitative data analysis, all 30 one-hour interviews were recorded, with automatic transcripts captured with the Zoom Captions function. As a pre-processing step, we manually checked the transcript to correct errors resulting from automatic transcription. The researchers then extracted \textit{all}’ verbalized quotes of all users that we can make sense of. To start with, we separate longer utterances into individual chunks of sentences that contain a single, clear idea, to distill fine-grained user attitudes without missing information accidentally when multiple ideas and preferences are bundled together. We created virtual sticky notes using an online tool named Miro, copied and pasted the individual utterances into them. These raw teacher quotes become the lowest level of affinity diagramming clustering. The researchers then grouped statements that have similar or recurring ideas into clusters, through an iterative affinity diagramming process \cite{Lucero2015-na}.
    
	Methodologically, we conducted three rounds of clustering: idea level (first level), challenge level (second level), and stakeholder level (third level) and to understand teachers and students’ preferences on level of control. We initially performed a clustering for user’ attitudes and reasoning toward \textbf{\textit{each design idea}} for all ideas under the seven design challenges. We then identified common themes within \textbf{\textit{each of the seven design challenges}} (described in section 4.1). Finally, we looked through the common themes identified within each design challenge to identify the main values, attitudes, considerations and concerns for \textbf{\textit{each stakeholder (teacher and student)}}. 
    
	To ensure reliability, three primary coders independently clustered data at the idea level using separate Miro boards. After the first round, coders met to review results and resolve discrepancies through workshop discussions to reach consensus on extracted themes, supervised by two faculty members through weekly (occasionally bi-weekly) synthesis meetings. For student data, we identified 114 first-level, 35 second-level, and 7 third-level clusters; for teachers, 148 first-level, 41 second-level, and 8 third-level clusters. Each affinity diagram—one for students and one for teachers—includes 15 large clusters used to compare preferences for classroom control. 
    
	For the rankings, we calculated the mean and standard deviation for a total of 34 ideas for the seven challenges. We then ranked the ideas for each challenge based on their mean scores, from the most favored (lowest mean score) to the least favored (highest mean score). To assess the level of agreement among teachers or students, we also calculated Kendall’s W \cite{Lumley2000-ca}, which has a range of 0 to 1. If the value of Kendall’s W is closer to 1, it indicates that the group of rankers is more unanimous and unified in their rankings within a challenge. Qualitative and quantitative analyses were combined and triangulated to interpret the most and least favorable design ideas for students and teachers regarding various challenges, with the goal of understanding their preferences for control distribution in the classroom orchestration process. Among the ideas evaluated, we mapped 32 design ideas into the design space (two design ideas that were not relevant to teacher-student hybrid control, thus we did not map it in the design space).

    \section{RESULTS}
    This section presents teacher and student preferences on classroom management, control, and agency in three key stages: before pairing, during collaboration, and after collaboration. Guided by the seven design challenges introduced earlier, we report both quantitative results (ranking data and Kendall’s W) and qualitative themes for each challenge to uncover their respective perspectives and control dynamics.
    \subsection{Stage 1: Before Pairing Up}
    \subsubsection{Challenge 1: Preparing Upcoming Collaboration Process Before Class.} Ensuring students are adequately prepared for peer interaction or tutoring at the start of class is crucial. This presents challenge 1 \textit{“How might we design a tool to help teachers prepare for the upcoming collaboration process before class?”} Researchers, teachers, and students co-designed 5 ideas to solve this challenge: 1) Pairing rules before class, 2) Quiz for pairing, 3) Students share preferences, 4) Warm-up topics, and 5) Personality test for pairing (Table \ref{Table 1}). 

We found that overall, teachers’ opinions are more unified, while students’ views on the design ideas related to challenge 1 are more varied 
(\textit{Kendall’s W (teachers)} \textnormal{= 0.533}, \textit{Kendall’s W (students)} \textnormal{= 0.284}). Based on the mean and SD in Table \ref{tab:challenge1}, students ranked “Quiz for pairing” as the most favorable idea, while teachers ranked “Pairing rules before class”, closely followed by “Quiz for pairing”. Students ranked lowest with the idea of “Warm-up topics”, closely followed by "Pairing rules before class" as the second least favored idea. The teachers view the idea of “Personality test for pairing” as their least favorable.
\begin{table}[!htbp]
  \centering
  \renewcommand{\arraystretch}{1.15}
  \caption{Design ideas and stakeholders rankings for \textbf{Challenge 1}: How might we prepare for the upcoming collaboration process before class? 1 = most preferred, 5 = least preferred}
  \label{tab:challenge1}
 \begin{tabularx}{\textwidth}{
    p{3cm}
    >{\raggedright\arraybackslash}X
    c c c c    % 改这里
}
    \toprule
    \textbf{Design Ideas} & \textbf{Idea Description} &
    \makecell{\textbf{Mean}\\Rank\\(Stu.)} &
    \makecell{\textbf{SD}\\(Stu.)} &
    \makecell{\textbf{Mean}\\Rank\\(Tea.)} &
    \makecell{\textbf{SD}\\(Tea.)} \\
    \midrule
    % ------- rows -------
    D1. Pairing rules before class &
      Teachers specify rules and criteria for pairing in the system.\newline
      Students can view these criteria before starting in-class collaboration. &
      3.7 & 1.19 & 2.5 & 1.31 \\

    D2. Quiz for pairing &
      Students take a quiz assessing their abilities;\newline
      teachers review the results and pair students accordingly. &
      1.7 & 1.01 & 2.7 & 1.4 \\

    D3. Students share preferences &
      Students tell teachers their partner preferences, and teachers consider them when pairing. &
      2.9 & 1.45 & 3.1 & 1.39 \\

    D4. Warm-up topics &
      Teachers give students small-talk topics so they can discover partners’ interests before collaborating. &
      3.8 & 1.33 & 2.8 & 1.33 \\

    D5. Personality test for pairing &
      The system offers a personality test; compatible partners are identified and teachers assign pairs accordingly. &
      2.9 & 0.95 & 3.4 & 1.36 \\
    \bottomrule
  \end{tabularx}
  \label{Table 1}
\end{table}

Students valued using \textbf{quizzes} and \textbf{personality tests}, as well as direct ways to \textbf{share preferences with teachers}, to help teachers and the system better understand their strengths and weaknesses and improve pairing. 
While some students felt that knowing the rules and criteria for pairing helped them focus or prepare in advance, 
many emphasized that having the right partner mattered more in collaborative contexts: 
\textit{“Getting a partner that works better is more important than knowing the rules”} (S5). 
Students had mixed feelings about warm-up topics: some thought they might be \textit{“potentially distrac[ting] and make them think of something off-topic”} (S2), while others believed they could help students \textit{“see what their partners are interested in”} (S6).

Teachers did not like the idea of using personality tests for pairing as much as students, expressing concerns about accuracy and practicality. 
One teacher noted, \textit{“I know my students, and I know they might like to give wrong answers [on the personality quiz] intentionally”} (T6), highlighting concerns about questionable data quality. Some doubted the scientific validity, claiming that they are \textit{“...not based on any science, not sure how useful this would be... This is more noise than useful information that you can pair with”} (T2). 
Consequently, teachers preferred data-driven methods such as quizzes, believing that they provide more accurate academic assessments to ensure \textit{“the tutor is going to be able to actually tutor the other students”} (T3) to reduce teacher workload during pairing. Although most teachers believed that knowledge level is a key factor when pairing students, there are \textit{“lots of different elements that go into successful collaboration”} (T15), such as personality compatibility, student preferences, social and emotional needs, or even cultures. 
In these cases, they rely on their experience and classroom management skills rather than standardized assessments for these factors.

\vspace{\baselineskip}

\noindent {\textbf{Stage 1 "Before Pairing Up" Summary}: Students focus less on knowing the specific rules and more on finding a good partner; they favor quizzes and personality tests to help the system understand their needs. Conversely, teachers prefer pairing rules and academic quizzes to ensure students are prepared to support one another. Teachers prioritize data-driven methods and academic levels, while students focus more on personal comfort and compatibility.}

\subsection{Stage 2: During Pairing Up}
\subsubsection{Challenge 2: Ensuring Collaboration Starts at a Proper Time for Both Teachers and Students} Collaboration activities initiated and decided by teachers may not take into account student readiness and preferred timing. 
Challenge 2 addresses the question \textit{“How can we design the tool to ensure that collaboration begins at an appropriate time?”}. 
Researchers, teachers, and students co-designed five ideas to solve this challenge: 
1) Teacher decides, 2) When students are ready, 3) Same time for everyone, 4) Students decide, 5) Students propose, teacher decides (Table \ref{Table 2}).

Based on Kendall’s coefficients, students and teachers showed a similar, moderate level of agreement in ranking these options 
(\textit{Kendall’s W (teachers)} \textnormal{= 0.533}, \textit{Kendall’s W (students)} \textnormal{= 0.55}). 
For students, based on the mean and SD in Table \ref{tab:challenge2}, “When students are ready” was the top choice, followed by “Students propose, teacher decides”. The least favored idea for students was “Students decide”. \textcolor{blue}{Teachers most preferred “When students are ready” as well, followed by “Teacher decides”.} Teachers ranked “Same time for everyone” lowest, closely followed by "Students decide".

\begin{table}[!htbp]
  \centering
  \renewcommand{\arraystretch}{1.15}
  \caption{Stakeholder rankings for \textbf{Challenge 2}: How can we ensure that collaboration begins at an appropriate time? 1 = most preferred, 5 = least preferred}
  \label{tab:challenge2}
 \begin{tabularx}{\textwidth}{
    p{3cm}
    >{\raggedright\arraybackslash}X
    c c c c    % 改这里
}
    \toprule
    \textbf{Ideas} & \textbf{Idea Description} &
    \makecell{\textbf{Mean}\\Rank\\(Stu.)} &
    \makecell{\textbf{SD}\\(Stu.)} &
    \makecell{\textbf{Mean}\\Rank\\(Tea.)} &
    \makecell{\textbf{SD}\\(Tea.)} \\
    \midrule
    D1. Teacher decides &
      Based on the teacher’s judgment, the teacher starts the collaboration with the students. &
      3.5 & 1.12 & 2.1 & 1.09 \\

    D2. When students are ready &
      Students have a \emph{“Ready to collaborate”} button; when toggled, the teacher can see and start the collaboration. &
      1.2 & 0.40 & 1.7 & 0.87 \\

    D3. Same time for everyone &
      Collaboration starts/ends simultaneously for the entire class. &
      3.1 & 0.95 & 3.8 & 1.23 \\

    D4. Students decide &
      Students start a collaboration themselves (without consulting teachers). &
      4.4 & 0.67 & 3.6 & 1.26 \\

    D5. Students propose, teacher decides &
      Students can request not to be paired and keep individual learning; teachers approve or deny the request. &
      2.8 & 1.33 & 2.7 & 1.34 \\
    \bottomrule
  \end{tabularx}
  \label{Table 2}
\end{table}

Teachers favored “When Students are ready” because \textit{“Sometimes students, especially those with accommodations or needing extra support, might not know when they’re ready”} (T6). 
They worried that letting students decide could lead to chaotic classroom management: 
\textit{“Teachers know what content students should collaborate. on, if students do whatever they want, it would be chaotic”} (T8) or 
\textit{“... might mess up teachers' schedules with too much freedom”} (T10).

Students, on the other hand, preferred “When students are ready” because it lets them start collaborating at their own pace. 
S2 noted: \textit{“Students can share when they are ready because some students take longer than others. If there is a ‘ready to collaborate’ button, they can press start whenever they are ready instead of rushing.”} 
Students appreciated a balance between autonomy and teacher guidance, seen in their second-choice “Students propose, teacher decides”: 
\textit{“Some might want to individually work to see if they can do it without someone. If the teacher feels like the student needs help, they can pair the students with another person”} (S7). 
Despite wanting more autonomy, students also expressed concerns about the inefficiency of “Students decide”. 
One student said: \textit{“If students collaborate whenever they want and choose their partners themselves, it might get chaotic, and teachers will not know if students will complete the work”} (S5). 
They also feared that self-selected partners might reduce productivity because \textit{“students might pick people they are friends with and end up never doing the work”} (S3).

In conclusion, the ideas that best address collaboration timing are “When students are ready” for both teachers and students. Both groups expressed concerns about giving students full control, including potential impacts on classroom discipline, as well as the risk of overlooking differences in students’ needs.

\subsubsection{Challenge 3: Assigning Suitable Roles and Partners that Satisfy Both Teachers and Students}

When only teachers choose partners and roles, students may feel less motivated and want more say. 
To address this, seven ideas were co-created for Challenge 3: 1) Teacher decides, 2) Teacher proposes, students decide, 3) Students propose, teacher decides, 4) Students decide,  
5) Teacher assigns, student chooses role, 6) Students are able to reject partners, and 7) Alternate choices (Table \ref{Table 3}).

For pairing decisions, teachers showed more uniform views than students (Kendall’s W: \textit{teachers} \textnormal{= 0.484}, \textit{students} \textnormal{= 0.284}). Based on the mean and SD in Table \ref{tab:challenge3}, students ranked “Teacher decides” highest and “Student decides” lowest, indicating a preference for structured guidance. Teachers preferred a balanced approach, ranking “Teacher assigns, student chooses role” top and “Students are able to reject partners” last. Both groups favored shared decision-making: “Teacher proposes, students decide” and “Students propose, teacher decides” were common second- and third-ranked ideas, showing a willingness to balance teacher authority with student input.

\begin{table}[!htbp]
  \centering
  \renewcommand{\arraystretch}{1.15}
  \caption{Design ideas and stakeholders rankings for \textbf{Challenge 3}: Assigning suitable roles and partners that satisfy both teachers and students. 1 = most preferred, 5 = least preferred}
  \label{tab:challenge3}
\begin{tabularx}{\textwidth}{
    p{3cm}
    >{\raggedright\arraybackslash}X
    c c c c    % 改这里
}
    \toprule
    \textbf{Design Ideas} & \textbf{Idea Description} &
    \makecell{\textbf{Mean}\\Rank\\(Stu.)} &
    \makecell{\textbf{SD}\\(Stu.)} &
    \makecell{\textbf{Mean}\\Rank\\(Tea.)} &
    \makecell{\textbf{SD}\\(Tea.)} \\
    \midrule
    D1. Teacher decides & Teachers choose partners for students based on their past collaboration and skill information.  & 1.9 & 1.52 & 3.13 & 1.59 \\
    D2. Teacher proposes, students decide & Teacher specifies three partners; students pick one. & 4.0 & 1.49 & 2.74 & 1.48 \\
    D3. Students propose, teacher decides & Students list top-3 partners; teacher picks one. & 3.2 & 1.40 & 2.67 & 1.5 \\
    D4. Students decide & Students choose partners without teachers. & 5.4 & 1.75 & 4.07 & 1.53 \\
    D5. Teacher assigns, student chooses role & Teachers assign partners; students pick role. & 4.5 & 1.37 & 2.2 & 1.17 \\
    D6. Students can reject partners & Students may reject and switch partners. & 4.8 & 1.73 & 5.2 & 1.98 \\
    D7. Alternate choices & One class teacher pairs, next class students choose. & 4.2 & 2.40 & 4.3 & 2.19 \\
    \bottomrule
  \end{tabularx}
  \label{Table 3}
\end{table}

Based on the qualitative data, both teachers and students preferred that teachers control pairings while allowing students some autonomy in selecting roles. Students supported teacher-led control for three reasons: 1) teachers know students’ skills and history, 2) they can better identify suitable partners, and 3) concerns about friends causing distraction. For example, they appreciated teachers preventing situations where students might \textit{“...choose their friends, resulting in potentially ineffective collaboration”} (S10). Still, some valued choice for familiar partners and higher engagement, as one said, \textit{“You might not be very productive if you don’t like the partner you are working with”} (S4).

Teachers likewise valued guiding pairings while allowing limited student input to enhance learning. This aligns with their priorities of managing classroom dynamics effectively, ensuring appropriate pairings while giving students some input into their roles. While many teachers believed that \textit{“teachers can make better choices based on class settings”} and that \textit{“teacher decisions would be more efficient,”} they also recognized the importance of understanding students’ preferences by involving students in decisions to increase learning motivation (T9). Generally, teachers agreed that allowing students some freedom, such as proposing partner choices, fosters greater motivation and engagement.

At the same time, teachers worried that too much freedom might reduce productivity, citing risks that 1) students take too long to decide, 2) lack information about peers, and 3) add to teacher workload in managing poor decisions.

\subsubsection{Challenge 4: Selecting Suitable Content for Students that Satisfy Both Teachers and Students.} 

Prior work indicated that if the collaborative topics and content are determined solely by teachers, it might limit students’ autonomy, potentially decreasing their engagement. This reveals the \textit{“How might we design the tool to select suitable collaboration content for students to work on?”} Five ideas were co-designed to address this challenge:  
1) Teacher choice, 2) Teacher proposes, students decide, 3) Students propose, teacher decides, 4) Students can switch tasks, and 5) Students choice (Table \ref{Table 4}).

Overall, based on Kendall’s coefficients, teachers had a higher agreement among ranker for challenge 4 than students (\textit{Kendall’s W teachers} \textnormal{= 0.559}; \textit{Kendall’s W students} \textnormal{= 0.496}), suggesting that teachers had more consistent opinions. Based on the mean and SD in Table \ref{tab:challenge4}, teachers and students had different preferences in ideas but similar in attitudes regarding the decision of collaboration content: teachers preferred “Teacher proposes, students decide,” while students favored “Students propose, teacher decides.” Both ideas are on the hybrid-control spectrum. Both groups ranked “Student choice” as their least preferred option.

\begin{table}[H]
  \centering
  \renewcommand{\arraystretch}{1.15}
  \caption{Design ideas and stakeholders rankings for \textbf{Challenge 4}: Selecting suitable collaboration content. 1 = most preferred, 5 = least preferred}
  \label{tab:challenge4}
 \begin{tabularx}{\textwidth}{
    p{3cm}
    >{\raggedright\arraybackslash}X
    c c c c    % 改这里
}
    \toprule
    \textbf{Ideas} & \textbf{Idea Description} &
    \makecell{\textbf{Mean}\\Rank\\(Stu.)} &
    \makecell{\textbf{SD}\\(Stu.)} &
    \makecell{\textbf{Mean}\\Rank\\(Tea.)} &
    \makecell{\textbf{SD}\\(Tea.)} \\
    \midrule
    D1. Teacher choice & Teachers decide what content students work on. & 3.1 & 1.38 & 2.5 & 1.63 \\
    D2. Teacher proposes, students decide & Teacher proposes 3 tasks; students pick one. & 2.0 & 1.10 & 1.6 & 0.62 \\
    D3. Students propose, teacher decides & Students propose 3 tasks; teacher chooses. & 1.83 & 0.38 & 2.3 & 0.61 \\
    D4. Student can switch task & Students may switch topics if dissatisfied. & 3.3 & 1.19 & 3.5 & 1.15 \\
    D5. Student choice & Students decide their own collaborative tasks. & 3.5 & 1.29 & 3.7 & 0.95 \\
    \bottomrule
  \end{tabularx}
  \label{Table 4}
\end{table}

There are five key benefits to teachers selecting collaboration content: assigning tasks that matches well with students’ abilities, saving time, ensuring the logical progression of skills development, meeting educational standards, and reducing disruptions. While teachers generally believe they should select the content, they also valued student involvement when:  
1) teachers provide structured guidelines on content choices,  
2) tasks are lighter or more engaging, and  
3) students could gauge their own abilities.
As one teacher explained, \textit{“We might have a broad topic, and then they pick subtopics underneath...”} (T6).  
Teachers believe the “Teacher proposes, students decide” approach offers flexibility, increasing student engagement while ensuring content aligns with required standards. In the mean time, students felt they best understood their own difficulties and preferred proposing topics, while relying on teachers to adjust content difficulty as needed. 

Both groups recognized risks of “Student choice”: students might avoid challenges or lack maturity to select appropriately. 
As one teacher noted, \textit{“It’s teachers’ responsibility to make sure that [student choice] idea is mature”} (T1). Another added, \textit{“They might pick the thing that has the least amount of work when they could handle more”} (T14).

In sum, the preferred approach for selecting collaboration content is “Teacher proposes, students decide” for teachers and “Students propose, teacher decides” for students.  
The choice between these approaches should be adapted to classroom circumstances, ensuring the content meets the needs of both groups.

\subsubsection{Challenge 5: Ensuring Productive Collaboration.} For Challenge 5, \textit{“How might we design the tool to ensure the collaboration is productive?”}, researchers, teachers, and students co-designed four ideas:  
1) System detects keywords, 2) Request to change assignment, 3) Request to change partner, and 4) Get help from NPC (details in Table \ref{tab:challenge5}). In this challenge, ideas involving automated supports were proposed as a way to facilitate classroom control distribution.

Based on Kendall’s coefficients, teachers had higher agreement among the rankers for Challenge 5 than the students (\textit{Kendall’s W teachers} \textnormal{= 0.748}; \textit{Kendall’s W students} \textnormal{= 0.108}).  
From the mean and SD in Table \ref{tab:challenge5}, students preferred “Get help from NPC” and ranked it as the highest, reflecting strong interest in utilizing automated support to monitor collaboration and provide assistance. Teachers ranked "System detects keywords" as the highest, closely followed by "Get help from NPC", suggesting teachers' needs on having additional help on supervising.
“Request to change assignment” was the least favored by both groups, closely followed by “Request to change partner”.
\begin{table}[H]
  \centering
  \renewcommand{\arraystretch}{1.15}
  \caption{Design ideas and stakeholders rankings for \textbf{Challenge 5}: How might we ensure the collaboration to be productive? 1 = most preferred, 5 = least preferred}
  \label{tab:challenge5}
  \begin{tabularx}{\textwidth}{
      p{3cm}
      >{\raggedright\arraybackslash}X
      cccc
  }
    \toprule
    \textbf{Ideas} & \textbf{Idea Description} &
    \makecell{\textbf{Mean}\\Rank\\(Stu.)} &
    \makecell{\textbf{SD}\\(Stu.)} &
    \makecell{\textbf{Mean}\\Rank\\(Tea.)} &
    \makecell{\textbf{SD}\\(Tea.)} \\
    \midrule
    D1. System detects keywords &
      System flags disengagement words (e.g.\ “video game”) and alerts the teacher. &
      2.3 & 1.27 & 1.5 & 0.72 \\

    D2. Request to change assignment &
      Students can complain about the assignment and request a change. &
      2.9 & 0.84 & 3.3 & 0.60 \\

    D3. Request to change partner &
      Students can complain about partners and request switching, unseen by the partner. &
      2.8 & 1.08 & 3.1 & 0.93 \\

    D4. Get help from NPC &
      Students press a button to get NPC help; if unresolved, teacher steps in. &
      2.0 & 1.00 & 1.7 & 0.79 \\
    \bottomrule
  \end{tabularx}
  \label{Table 5}
\end{table}

According to qualitative data, teachers, who often lack the time to help all struggling students, appreciated NPCs as a way to streamline support, especially for those who \textit{“don’t want to ask for help or are shy about asking”} (T6). Students also viewed NPCs as a useful backup when teachers or peers were unavailable. However, concerns emerged about NPC effectiveness, with teachers questioning their “smartness” and students fearing that some might \textit{“take advantage and keep asking the NPC for help”} (S4). Both groups also favored “System detects keywords” as a lightweight monitoring tool, indicating demand for support that reduces teacher workload while keeping students on track.

Ideas involving requests to change partners or assignments ranked lowest, as the drawbacks outweighed potential benefits. Students worried such options could lead to easier tasks, overuse, or harm to relationships. Teachers were even more opposed, emphasizing that well-structured courses and carefully chosen partners reduce the need for changes. As one teacher explained, \textit{“I try to avoid this option as teachers put a lot of effort and time into designing the assignment”} (T11). Another noted, \textit{“Students should learn to work with different people, which can improve their social skills”} (T10). While both groups agreed that requests for changes shouldn’t be completely prohibited, teachers were cautious, whereas students were somewhat more open to justified cases. Overall, changes were considered a last resort.

In sum, both teachers and students favored automated support to sustain productive collaboration, especially when teachers were unavailable, which reflects a dynamic distribution of control where part of it is delegated to external tools. However, both groups expressed reservations about overusing students' requests to change assignments or partners, viewing them as options to be used rarely due to concerns about classroom efficiency and the potential disruption to social learning.

\vspace{\baselineskip}

\noindent {\textbf{Stage 2 Summary: During Pairing (C2, C3, C4)}: For pairing timing (C2), both teachers and students favor transitioning only when students are ready, accommodating different student paces. Teachers have a stronger preference for starting all students at the same time for easier classroom management, while students want to voice their preferred mode. For partner choice (C3), students favor “Teacher decides,” while teachers prefer a balanced option, which is “Teacher assigns, student chooses role.”  Teachers support partial student autonomy but not full control (e.g. rejecting partners). Both groups value shared decision-making (e.g., “Teacher proposes, students decide”). For collaboration content (C4), both groups agreed it is least favorable to leave decisions solely to students, fearing they may choose easier topics. Our study suggests that adaptivity should depend on task seriousness: teachers prefer selecting certain content to ensure standards, logical progression, and management, but are more open to student agency when tasks are less demanding.
{\textbf{During Collaboration (C5).} Both teachers and students emphasized the value of automated support to facilitate collaboration (e.g., “Get help from NPC”). Teachers and students both agreed that requests to change assignments or partners should be a last resort, only allowed in justified cases to avoid disrupting learning and social dynamics.}}

\subsection{Stage 3: After Pairing Up}

\subsubsection{Challenge 6: Facilitating Smooth Transitions Between Collaborative and Individual Activities}

Ensuring smooth transitions between individual and collaborative learning activities is challenging without disrupting students’ learning continuity and quality. Challenge 6 focuses on \textit{“How might we design the tool to ensure students have a smooth transition between collaborative and individual activities?”}. Four ideas were co-designed to address this:  
1) Stop when teacher asks,  
2) Join now/join later,  
3) Countdown time/problems, and  
4) Save progress and resume (Table \ref{tab:challenge6}).
\begin{table}[!htbp]
  \centering
  \renewcommand{\arraystretch}{1.15}
  \caption{Design ideas and stakeholders rankings for \textbf{Challenge 6}: How might we ensure students have a smooth transition between collaborative and individual activities? 1 = most preferred, 5 = least preferred}
  \label{tab:challenge6}
  \begin{tabularx}{\textwidth}{
      p{3cm}
      >{\raggedright\arraybackslash}X
      cccc
  }
    \toprule
    \textbf{Ideas} & \textbf{Idea Description} &
    \makecell{\textbf{Mean}\\Rank\\(Stu.)} &
    \makecell{\textbf{SD}\\(Stu.)} &
    \makecell{\textbf{Mean}\\Rank\\(Tea.)} &
    \makecell{\textbf{SD}\\(Tea.)} \\
    \midrule
    D1. Stop when teacher asks &
      When a teacher chooses to “unpair” students, collaboration stops immediately. &
      3.6 & 0.49 & 3.5 & 0.72 \\

    D2. Join now / join later &
      Students may finish current individual problems and join collaboration later. &
      2.2 & 1.17 & 2.7 & 0.93 \\

    D3. Count-down time / problems &
      Teacher displays remaining questions or time to end collaboration. &
      2.6 & 1.02 & 1.8 & 0.55 \\

    D4. Save progress and resume &
      System saves unfinished progress; students resume on their own time. &
      1.6 & 0.49 & 1.3 & 0.70 \\
    \bottomrule
  \end{tabularx}
  \label{Table 6}
\end{table}

Overall, based on Kendall’s coefficients, students had a higher agreement than teachers (\textit{Kendall’s W teachers} \textnormal{= 0.424}, \textit{Kendall’s W students} \textnormal{= 0.702}). Based on the mean and SD in Table \ref{tab:challenge6}, Both teachers and students favored “Save progress and resume” as the top choice and placed “Stop when teacher asks” last. "Join now / join later" and "Count-down time / problems", two hybrid control ideas, were ranked in the middle, with teachers slightly favoring "Count-down time / problems" than students and vice versa.

This suggests both teachers and students viewed abrupt transitions as disruptive. Immediate stopping interrupts students mid-task and adds pressure, whereas saving progress allows for flexible pacing and continuity. Qualitative results also validates this. As one student noted, \textit{“If students are in the middle of something, they can save their work and next time they collaborate, they don’t need to restart”} (S5). Teachers echoed this, stating that saving progress reduces the need to rush and better supports learning.

In the case of “Stop when teacher asks,” both students and teachers found it disruptive. Students suggested that unpairing should be gradual, with a warning or extra time to wrap up discussions, as \textit{“teachers should give students a warning to let them know what’s happening”} (S9). Teachers noted that an abrupt stop could hinder collaboration, as \textit{“it doesn’t give them enough time to collaborate properly”} (T4). Overall, both teachers and students favor “Save progress and resume” as the best solution for ensuring a smooth transition between collaborative and individual activities, while “Stop when teacher asks” was viewed as too abrupt. Providing advance warnings and allowing students to finish their discussions would further support smoother transitions.

\subsubsection{Challenge 7: Evaluating Collaboration Activities for Effective Future Pairing}

For Challenge 7 \textit{“How might we design the dynamic co-orchestration tool to effectively evaluate collaboration activity, ensuring good pairing to occur again in the future?”}, four ideas were co-generated to solve it:  
1) Teacher pairs based on collaboration data/statistics,  
2) Teacher pairs based on rating,  
3) Both teachers and students rate,  
4) Students rate (Table \ref{tab:challenge7}).

Overall, based on Kendall’s coefficients, teachers had a higher agreement among all rankers for Challenge 7 than students (\textit{Kendall’s W (teachers)} \textnormal{= 0.693}, \textit{Kendall’s W (students)} \textnormal{= 0.410}). Based on the mean and SD in Table \ref{tab:challenge7}, for students, they ranked “Teacher pairs based on collaboration data”, an idea leaning towards teacher control, as their top choice. Teachers ranked “Both teachers and students rate”, an idea leaning towards hybrid control, as their favorite idea. For both groups, “Students rate” was their least favored idea.

\begin{table}[!htbp]
  \centering
  \renewcommand{\arraystretch}{1.15}
  \caption{Design ideas and stakeholders rankings for \textbf{Challenge 7}: How might we evaluate collaboration activity so good pairing can occur again in the future? 1 = most preferred, 5 = least preferred}
  \label{tab:challenge7}
  \begin{tabularx}{\textwidth}{
      p{3cm}
      >{\raggedright\arraybackslash}X
      cccc
  }
    \toprule
    \textbf{Ideas} & \textbf{Idea Description} &
    \makecell{\textbf{Mean}\\Rank\\(Stu.)} &
    \makecell{\textbf{SD}\\(Stu.)} &
    \makecell{\textbf{Mean}\\Rank\\(Tea.)} &
    \makecell{\textbf{SD}\\(Tea.)} \\
    \midrule
    D1. Teacher pairs based on data &
      Using collaboration statistics (e.g.\ problems finished, correctness), teacher forms future pairs. &
      1.4 & 0.49 & 2.1 & 0.96 \\

    D2. Teacher pairs based on rating &
      After collaboration, students rate partners and activity; teacher uses ratings for future pairing. &
      2.9 & 0.84 & 2.6 & 1.02 \\

    D3. Both teachers \& students rate &
      Both parties rate the collaboration; system uses ratings to form pairs. &
      2.3 & 1.10 & 1.5 & 0.5 \\

    D4. Students rate &
      Students rate the collaboration; system stores ratings for future pairing. &
      3.3 & 0.79 & 3.4 & 0.88 \\
    \bottomrule
  \end{tabularx}
  \label{Challenge 7 Ideas}
\end{table}

Both teachers and students recognize the value of system-generated data, as it provides objective insights and reduces bias in forming collaborative pairs. As one teacher explained, \textit{“It gives teachers and students a chance to provide ratings, and the system can use that information to form future pairs”} (T15). Students favor teacher-led pairings based on collaboration data, trusting that teachers, supported by the system, can form logical and effective pairs by assessing how well the collaboration went. One student noted, \textit{“It can give teachers ideas and statistics to pair them again in the future and help them become more efficient”} (S1). Students also believe their ratings offer valuable insights into partner performance and experiences that system data may not capture. Teachers agree that recognizing this input is the key to creating more effective pairings. As one teacher shared, \textit{“Some students need more help, and I might want to pair them with stronger students”} (T6). However, teachers prefer to make the final decisions, saying, \textit{“I like to hear from my students, but ultimately I prefer to make the decision”} (T3).

Despite the value of student ratings, both groups are cautious about potential biases in relying solely on these ratings for future collaboration decisions. As one student pointed out, \textit{“Students might rate their friends really high to be paired again, even if they won’t be productive”} (S7). Teachers echoed this concern, fearing that \textit{“students might give low ratings to avoid working with certain peers”} (T14). Nonetheless, both agree that students’ subjective experiences should be considered alongside data. In conclusion, while students favor “Teacher pairs based on collaboration data/statistics” for future pairings, teachers prefer a balanced approach, using both student ratings and system data to make informed decisions.

\vspace{\baselineskip}

\noindent {\textbf{Stage 3 Summary: After Pairing (C6, C7)}: For transitions between learning modes (C6), teachers and students strongly favored allowing students to save progress and resume, supporting continuity, self-paced work, and reduced stress. In contrast, they both don’t like sole teacher control and for students to immediately stop collaboration when teachers ask due to its abruptness and disruption. Both groups agreed that advance warning and time to wrap up would further ease transitions. For evaluating collaborative learning activity for the future (C7), teachers and students valued system-generated data to inform pairing, as it provides objective insights and reduces bias. Students ultimately trust teachers, with data support, to decide, while teachers prefer balancing system data with student input.}

\section{OVERALL DISCUSSION}
\subsection{Stakeholders Preferences that Align and Differ}

\begin{figure}
    \centering
    \includegraphics[width=\linewidth]{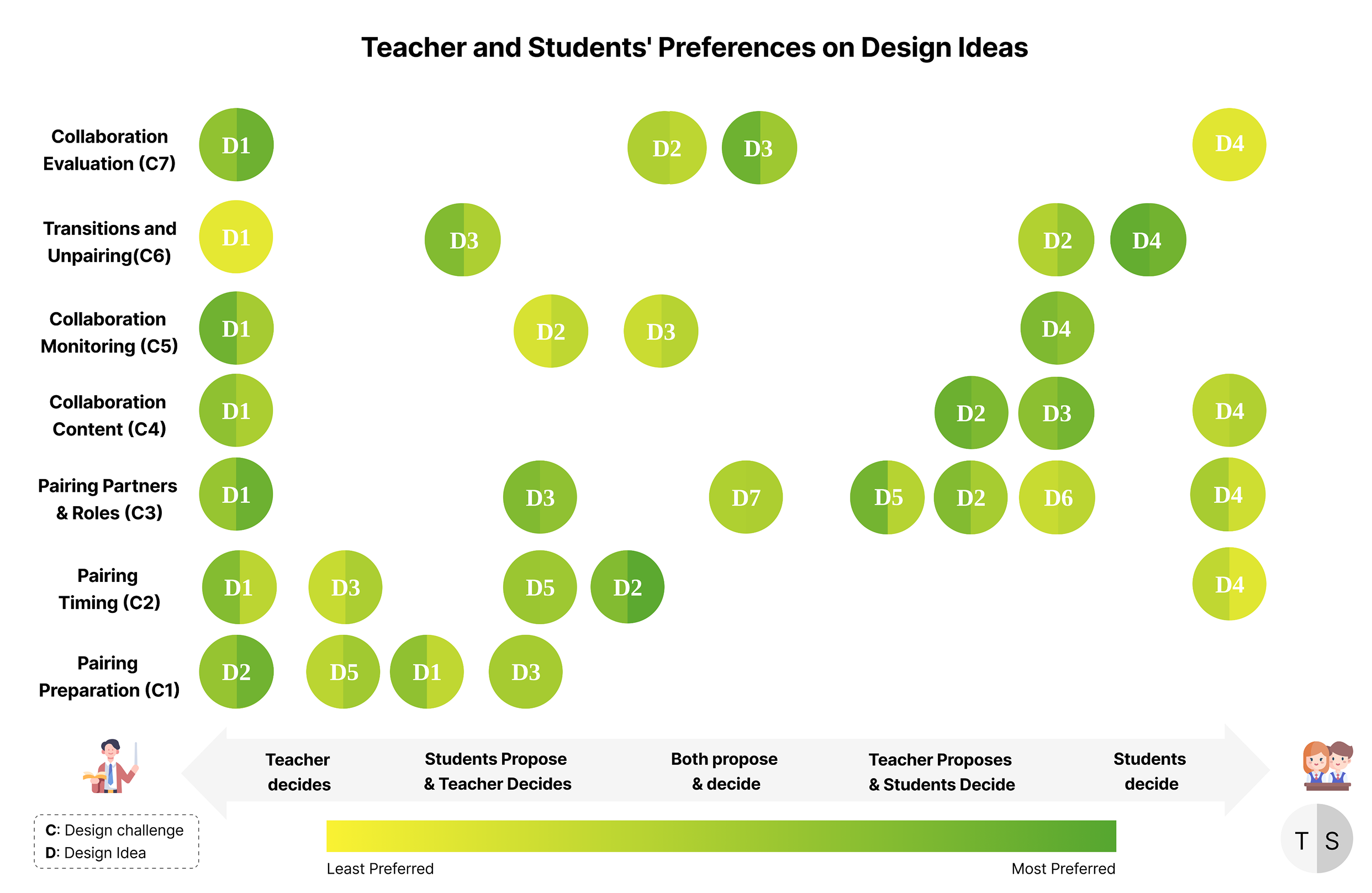}
    \caption{Visualization of stakeholder preferences across design challenges (C1-C7). The horizontal axis represents teacher-student agency levels, while the vertical axis maps user journey stages. Color-coded bubbles show teacher (left) and student (right) preferences, with greener shades indicating favorable ideas and yellow shades indicating less favorable ones. Stronger contrasts highlight differing opinions.}
    \label{fig:Design_Space}
\end{figure}

\textcolor{blue}{We aggregate teachers’ and students’ preferences for design ideas across challenges by mapping each stakeholder group’s rankings to a yellow-to-green color scale shown at the bottom of the Fig. \ref{fig:Design_Space}. The horizontal axis depicts the spectrum of agency, ranging from teacher-dominated control on the left to student-dominated control on the right. The vertical axis maps the seven design challenges (C1--C7) corresponding to different stages in the co-orchestration journey—from pre-collaboration (bottom) to post-collaboration (top). Each design idea (labeled D) is represented as a color-coded bubble. Greener colors indicate more preferred ideas, whereas yellower colors indicate less preferred ones. This visualization allows us to quickly assess alignment between teachers and students (similar colors within a bubble) as well as divergence in their preferences (higher color contrast within a bubble).}

\textbf{Aligned Preferences.} Judging from the most favored ideas (dark green bubbles), both teachers and students preferred some form of \textbf{hybrid control} for collaboration content (C4), collaboration monitoring (C5), transitions and unpairing (C6), and collaboration evaluation (C7) (Fig. \ref{fig:Design_Space}; C4-D2; C4-D3; C5-D4; C6-D4; C7-D3). In contrast, both groups favored stronger \textbf{teacher control} during preparation prior to pairing (C1-D2). Additionally, both teachers and students expressed reservations about granting \textbf{students sole control} over pairing timing, partner selection, collaboration content, and collaboration evaluation, as indicated by the light yellow bubbles in the rightmost column of Fig. \ref{Fig. 14.}.

Interestingly, these results reveal a temporal pattern in which student involvement and agency are increasingly valued during mid- and post-activity stages, as the activity unfolds (Fig. \ref{fig:Design_Space}: C4-D2; C4-D3; C5-D4; C6-D4). One possible explanation is that, at these stages, students’ voices carry greater weight because they have accumulated direct experience with the task. Their feedback is therefore more likely to be grounded in lived positive or negative learning experiences, rather than in assumptions, biases, or attempts to game the system. Feedback at this point can help teachers flexibly adjust ongoing or future arrangements to better support students’ learning, aligning with the goal of more personalized and dynamically differentiated instruction.

At the same time, comparing the leftmost (greener) and rightmost (more yellow) bubbles suggests that both teachers and students generally find \textbf{\textit{full teacher control}} more acceptable than \textbf{\textit{full student control}} in classroom decision-making, In particular, allowing students to \textbf{request changes to the educational decision-making process} was often perceived as controversial and undesirable. Although such mechanisms afford greater student agency, stakeholders expressed concerns about potential misuse, such as repeatedly switching to easier tasks or selecting partners primarily for social rather than learning purposes.

\textbf{Differed Preferences.} \textbf{Teachers and students differ in their preferences for hybrid control over when to form pairs (C2) and who to pair with (C3).} Teachers preferred more control over pairing timing because they believed it helped maintain order and reduce chaos (C2-D1), while students favored greater agency to stay engaged (C3-D3, D5, D2). This aligns with previous findings on the downside of teachers taking sole control over pairing timing: students may feel unprepared or interrupted when they are orchestrated into a particular activity \cite{Yang2023-lq}. On the other hand, for deciding who to pair with whom, teachers desire more student voice since students may have interpersonal relationships or other pairing preferences that teachers are not aware of (C3-D5), while students have greater trust in their teacher’s choice rather than their own, thinking that teachers may have better judgment than students, especially from a pedagogical and curriculum planning standpoint (C3-D1).

\subsection{Design Implications for Mediating Teacher–Student Agency in Future Orchestration Tools}
\noindent{The findings from this study yield design implications about balancing teacher-student agency, for future teacher-facing orchestration systems in collaborative learning settings. While teachers and students align on the value of smooth transitions and collaboration, they diverge on who should initiate and decide key orchestration actions and when agency should shift. This highlights the need for tools that move beyond single-stakeholder control toward hybrid, mediated agency. This aligned with prior work that conceptualizes teacher and student control as independent dimensions rather than a zero-sum trade-off  \cite{Eshel2003-vv}. Specifically, we outline several design recommendations grounded in our empirical findings.}

\textbf{Support student awareness and preparedness for orchestration actions.}
Aligned with prior work \cite{Prieto2019-orla}, we suggest orchestration tools to support teachers in clearly and intentionally communicating to students what to expect in the upcoming activities so that students have awareness. This includes when learning tasks will begin or end, what assignments are expected, and potential collaboration arrangements. As the collaboration learning activities progress, hybrid control (as compared to teacher control) may be more helpful, and it'd become more preferable for students to provide input for preferred timing (C2), roles (C3), and content (C4). By making upcoming orchestration actions visible in advance in learning activities such as previews, advance notices, or countdowns, tools can reduce surprise and disruption while affording students a basic form of agency grounded in anticipation rather than control. Prior work on orchestration has largely focused on supporting \textit{teacher awareness}, such as using multimodal sensors to generate orchestration graphs that help teachers visualize classroom states \cite{Prieto2018-dy, Prieto2019-orla}. Orchestration tool design, however, should also extend visibility and awareness about activities to students, which can cultivate a sense of agency and preparedness. This is especially crucial for orchestration tools that provide more personalized learning, where learning activities may start and end in a more dynamically differentiated manner for each student, leveraging various educational technology \cite{Dillenbourg2010-do,Molenaar2019-do,Morgan2014-gy,Tomlinson2014-gv,Yang2023-lq,Zhai2024-ac}.

\textbf{Enable structured and bounded student proposals for collaboration validated by teacher-facing analytics.}
Our findings suggest that student input is most valuable for pairing-related decisions where teachers may lack situational or interpersonal context. Yet, both stakeholders express concern about unrestricted student-initiated changes mid-activities, especially for requesting changes in partners or assignments (C5-D2, D3), which can be considered very disruptive. We therefore recommend future orchestration tools to allow students to submit structured proposals prior to collaboration (e.g., partner constraints, readiness signals, or role preferences) within teacher-defined guardrails, paired with analytics that situate these proposals relative to instructional goals (e.g., learning progress, compatibility indicators, or collaboration quality signals). In doing so, the tool provides a predictable resolution path when preferences conflict—students can be heard through structured input, while teachers retain constraints and override capacity to preserve manageability and pedagogical alignment.
These design recommendations aligns with prior work on learner models and learning analytics dashboards that support learner participation while enabling informed instructional decision-making. Specifically, it positions students as active agents/designers, supporting both teachers and students to influence the learning environment, to maximize students' academic achievement \cite{Eshel2003-vv, Dimitriadis2021-hcd}. Furthermore, analytics paired with bounded student proposals can situate input relative to instructional goals, supporting differentiation strategies and adjusting for individual students’ progress \cite{Molenaar2019-do, Morgan2014-gy, Wubbels2012-tsr}. Future orchestration tools should provide ways for students to express their preferences and reasoning informs teacher decisions, while support a hybrid regulation model (e.g., grounded in past collaboration analytics) to preserving pedagogical guardrails.

\textbf{Support stage-sensitive stakeholder agency across orchestration workflows.}
In our study, we observed that stakeholder preferences were not static, but varied across different orchestration stages: both groups preferred stronger teacher control during early preparation and initiation, while valuing more meaningful student input during and after collaboration, when students’ feedback is more likely to be grounded in lived activity experience, rather than presumptions or bias.

\textcolor{blue}{\textbf{Limitations.} We acknowledge that, despite our efforts to represent a broad spectrum of ideas, certain regions of the design space—such as \textbf{C1 with full student control}—remain unmapped. We encourage future work to further \textbf{expand and refine} this space. Furthermore, in the CSCW literature, existing concepts, such as \textbf{community-wide deliberation} prior to virtual team formation—where participants engage in structured discussions to prepare for collaboration—could offer novel insights to how \textbf{pre-collaboration readiness} can be ensured in educational settings \cite{Wen2017-wt}.}

\section{CONCLUSION}

This study investigated teachers’ and students’ opinions and evaluation of 34 design ideas on seven design challenges, in the context of designing co-orchestration tools that dynamically combine individual and collaborative learning. Through mixed-methods analysis, this study contributes insights into teachers’ and students’ preferences regarding hybrid control. We contribute to users’ preferences visualized in the hybrid-control design space, and offer design recommendations for future educational tools to achieve a balanced, hybrid control and agency for multi-stakeholders in classrooms. 
In the CSCW and HCI domain, Harris et al. studied the six key themes in online group formation: group composition, self-presentation, assembly mechanism, recruitment, organizing structures, and group culture \cite{Harris2019-yf}, which notably align with elements of the design space we contribute in this work (Fig. \ref{fig:Design_Space}, vertical axis) . For instance, group composition corresponds to Challenge C3: Pairing Partners and Roles, while assembly mechanism parallels the criteria for dynamic pairing in our study. The empirical findings from this design space may also offer valuable implications for broader HCI and
CSCW research on group formation tools that require hybrid control and multi-stakeholder agency.

Overall, the findings indicate a stronger preference for teacher control before and during pairing actions, with more student agency during collaboration, transitions, and post-collaboration evaluations. Teachers and students align more closely on control over learning content and collaboration monitoring but differ in their preferences for pairing timing and partner selection. Findings suggest that future educational tools adopt hybrid-control options allowing students \textbf{to be aware of arrangements, voice preferences, and give feedback.} However, it recommends caution when enabling students to change teacher-set arrangements, suggesting that this should be limited to specific contexts. Full student control over learning activities or the ability to override teachers is discouraged, as it risks disrupting the learning process. Lastly, we recommend that hybrid-control tools be designed to \textbf{increase student agency levels as the learning activity progresses.} To the field of educational technology and HCI, this study contributes empirical insights within a theoretical design space, focusing on balancing multi-stakeholder agency for teacher-facing classroom orchestration tools.

\begin{acks}
We sincerely thank the teachers and students who participated in this study for their time, engagement, and invaluable contributions. This work was supported by the National Science Foundation (NSF) under Grant No. 1822861. Any opinions, findings, and conclusions or recommendations expressed in this material are those of the authors and do not necessarily reflect the views of the NSF.
\end{acks}

%\section{Background and Related Work} \label{sec:related}

% intentionally left blank
% intentionally left blank
%\section{Discussion} \label{sec:discussion}

%\section{Conclusion} \label{sec:conclusion}

\bibliographystyle{ACM-Reference-Format}
\bibliography{Ref}

\appendix

\end{document}